\documentclass[a4paper]{jpconf}
\usepackage{graphicx}
\begin{document}
\title{Quasistationary states in a quantum dot formed at the edge of a topological insulator by magnetic barriers with finite transparency}

\author{D V Khomitsky$^*$, E A Lavrukhina}

\address{Department of Physics, National Research Lobachevsky State University of Nizhny Novgorod, 603950 Gagarin Avenue 23, Nizhny Novgorod, Russian
Federation}

\ead{$^*$ khomitsky@phys.unn.ru}

\begin{abstract}
A model of quasistationary states is constructed for the one-dimensional edge states propagating along the edge of a two-dimensional topological insulator based on HgTe/CdTe quantum well in the presence of magnetic barriers with finite transparency. The lifetimes of these quasistationary states are found analytically and numerically via different approaches including the solution of the stationary Schr\"odinger equation with complex energy and the solution of the transmission problem for a double barrier structure. The results can serve as a guide for determining the parameters of magnetic barriers creating the quantum dots where the lifetimes for the broadened discrete levels are long enough for manipulation with their occupation numbers by external fields.

\end{abstract}

\section{Introduction}

Quasistationary states appear as solutions of the Schr\"odinger equation with complex energy and finite lifetimes and are faced in numerous problems of quantum mechanics, especially in collision theory \cite{LL3,Zeld}. The lifetime for a state confined between the barriers is relevant for many nonstationary problems where the state of the structure is manipulated by external field. A simple estimate can be made in a way that the required lifetime should be greater than the corresponding manipulation times since for shorter lifetimes the "leakage" of the wavefunction outside of the barrier regions will occur. The results for the lifetimes are known for a great variety of atomic and solid state systems \cite{Iogansen,Azbel,Dymnikov,DVbook}. The solid state systems are mainly the semiconductor heterostructures with finite effective mass of the carriers. Much less is known for the lifetimes of quasistationary states in topological insulators (TI) which are in the focus of condensed matter physics during the last two decades \cite{TI1,TI2}. Their essential feature is the presence of highly conductive edge states while the Fermi level in the bulk is located within the energy gap. For a number of systems demonstrating the TI behavior the energy dispersion of the edge states has a massless, linear in momentum dependence. They are called the Dirac - Weyl fermions and have a typical point in the spectrum where the dispersion branches cross. One of the difficulties for the progress in TI applications in device structures is the problem of creation of compact objects for the edge state confinement like quantum dots (QD) in semiconductors. Various models of QD in TI have been proposed, most of them attributing the application of magnetic barriers \cite{Timm2012,Dolcetto2013,jetp2016}. In our recent paper \cite{jetp2020} we have applied the model of a QD formed by magnetic barriers with finite height but with zero transparency stemming from an infinite width to study the effects of driving by periodic electric field and the associated times of QD level population switch or the times of the escape into continuum. These times of coherent manipulation can be estimated as short as 20-30 ps. It is thus of big interest to consider the model of QD formed by the barriers of finite transparency where the lifetime of a quasistationary state formed at a broadened discrete level in the QD will be finite. Our goal in the present paper is to derive this model and to obtain the associated lifetimes for various regions of key parameters. In Sec.2 we build the model of quasistationary states and calculate the lifetimes as solutions of the Schr\"odinger equation with complex energy. In Sec.3 we calculate the lifetimes within the resonance tunneling approach and compare them with the previous ones, discussing their dependence on model parameters. In Sec.4 we present our conclusions.

\section{Hamiltonian and quasistationary states}

In papers \cite{jetp2016,jetp2020} we have derived a model of the discrete and continuum states in a 1D artificial quantum dot formed by the application of magnetic barriers in the proximity of the 1D edge states of the 2D topological insulator based on the HgTe/CdTe quantum well:

\begin{equation}
H=A k_y \sigma_z
-M_1 f_1(y) \left( \sigma_x \cos \theta_1 +\sigma_y \sin \theta_1 \right)
-M_2 f_2(y) \left( \sigma_x \cos \theta_2 +\sigma_y \sin \theta_2 \right).
\label{ham}
\end{equation}

Here $A=\hbar v_F$ is the parameter proportional to the Fermi velocity which determines the spectrum $E=\pm A k_y$ of the edge states. They can be described as massless Dirac fermions propagating along $Oy$, and for typical HgTe/CdTe QW the parameter $A=360$ $\rm{meV \cdot nm}$ \cite{TI1,TI2}. The magnetic barriers are described by the amplitudes $M_{1,2}$ reflecting the exchange interaction strength between the edge states and the magnets in energy units, and the orientations of the barrier magnetization in the $(xy)$ plane is defined by angles $\theta_{1,2}$. The functions $f_{1,2}(y)$ describe the barrier spatial profiles along $Oy$. We consider a simple step-like profile of $f_{1,2}$ for each magnetic barrier having the width $d_{1,2}$. This profile may describe both non-transparent barriers with the width $d_{1,2} \to \infty$ as it was in \cite{jetp2016,jetp2020} and the case of transparent barriers with finite $d_{1,2}$ which is the subject of the present paper. The alignment of the barriers is shown in Fig.1 where the areas from left of the left barrier to the right of the right barrier are labeled by $1 \ldots 5$, and an example of their magnetization directions is shown by arrows corresponding to the case $\theta_1=\theta_2=0$. The area $-L/2 \le y \le L/2$ between the barriers is where the confinement may take place and the discrete levels of such artificial quantum dot are formed.

\begin{figure}[ht]
\includegraphics[width=105mm]{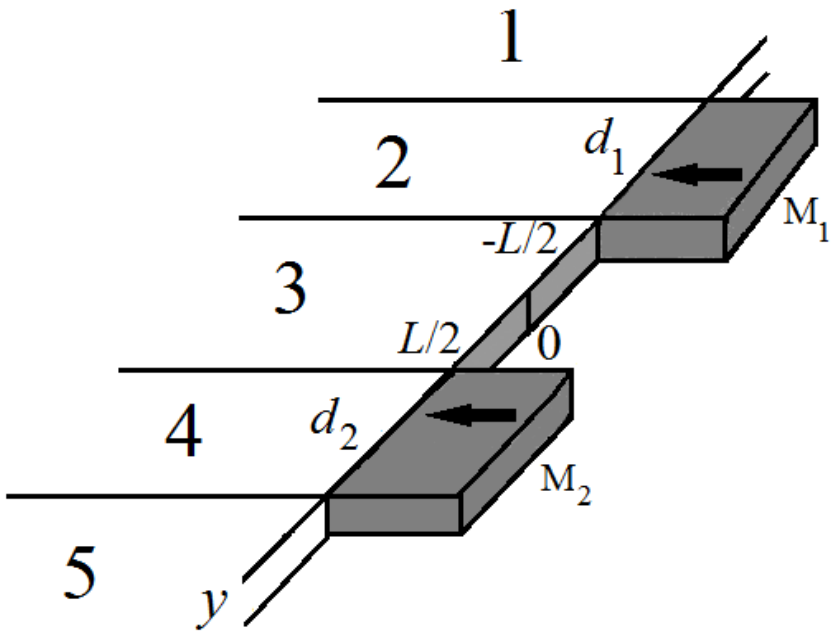}\hspace{1.3pc}%
\begin{minipage}[b]{50mm}\caption{\label{figint} The 1D edge (along $Oy$) of 2D topological insulator based on the HgTe/CdTe quantum well. Two magnetic barriers are attached at the edge with spacing $L$ between them. The barriers have the magnitude $M_{1,2}$ and width $d_{1,2}$, respectively. A variant of their magnetizations orientation $\theta_1=\theta_2=0$ is depicted by arrows. The areas labeled by 1..5 describe the spatial regions where the region 3 corresponds to the quantum dot between the barriers.}
\end{minipage}
\end{figure}

The solutions for the stationary Schr{\"o}dinger equation with Hamiltonian (\ref{ham}) may be written in a form of two-component spinors in the basis of the effective spin-1/2 for the edge states \cite{TI2}. In the areas outside the barriers (labeled as 1, 3, 5 in Fig.1) they can be expressed as

\begin{equation}
\psi^{\rm{out}}_{j}=
\left(
\begin{array}{l} 
C_{1j} \exp (iE_n y/A)  
\\ 
C_{2j} \exp (-iE_n y/A) 
\end{array}
\right),
\label{wf1}
\end{equation} 

where $j=1,3,5$. In the barrier regions $2, 4$ in Fig.1 the wavefunctions have the following form \cite{jetp2016,jetp2020}:

\begin{equation}
\psi^{\rm{barr}}_{j}=C_{1j} 
\left(
\begin{array}{l}  
1  \\ 
\beta_{1j}  
\end{array}
\right) 
\exp (-\gamma_j y)+
C_{2j} 
\left(
\begin{array}{l}
 1  \\ 
\beta_{2j}  
\end{array}
\right)
\exp (\gamma_j y),
\label{wf2}
\end{equation} 

where $j=2,4$. In Eq.(\ref{wf2}) the following notation is introduced:

\begin{equation}
\beta_{12,22}=\frac{\left(-E_n \pm i\sqrt{M_1^2-E_n^2}\right)
e^{i \theta_1}}{M_1},
\label{beta}
\end{equation}

\begin{equation}
\gamma_{2,4}=\frac{\sqrt{M_{1,2}^2-E_n^2}}{A},
\label{gamma}
\end{equation}

\begin{equation}
\beta_{14,24}=\frac{\left(-E_n \pm i\sqrt{M_2^2-E_n^2}\right)
e^{i \theta_2}}{M_2}.
\label{delta}
\end{equation}

We will consider a basic example of the energy spectrum under the approximation of two non-transparent barriers with $d_{1,2} \to \infty$ with two identical barriers having amplitudes $M_1=M_2=20$ meV and spacing $L=40$ nm between them. The magnetization orientation for both barriers is parallel with angles $\theta_1=\theta_2=0$. Under such conditions only two discrete levels are formed for the states confined in the quantum dot at $-L/2 \le y \le L/2$ \cite{jetp2020}. They have energies $E_{1,2}=\mp 9.62$ meV counted from the Dirac point $E=0$ of the spectrum. We will refer to this configuration below as a basic one also for the case of transparent barriers with finite width $d_{1,2}$.

The coefficients $C_{1j}$ and $C_{2j}$ in Eq.(\ref{wf1}),(\ref{wf2}) should be obtained from the boundary conditions. As in a problem of a quasistationary states for a particle with nonzero mass \cite{LL3,Zeld}, one may find a solution that has only outcoming waves in the outer regions 1 and 5 in Fig.1 but lacks the incoming ones. For the states (\ref{wf1}),(\ref{wf2}) this means the vanishing coefficients $C_{11}$ and $C_{25}$. The set of boundary conditions describes the continuity of the wavefunction at the interfaces between the regions 1-2, 2-3, 3-4, and 4-5 in Fig.1 which gives eight linear equations for ten coefficients $C_{1j}$ and $C_{2j}$, $j=1, \ldots, 5$. By keeping two of them as parameters, say, $C_{13}$ and $C_{33}$, we arrive at the following set of equations for vanishing coefficients $C_{11}$ and $C_{25}$ where the complex energy of a quasistationary state $\tilde{E}$ enters as a parameter:

\begin{equation}
\left\{
\begin{array}{l} 
C_{11}(C_{13},C_{23},\tilde{E})=0, \\ 
C_{25}(C_{13},C_{23},\tilde{E})=0.
\end{array}
\right.
\label{eqzero}
\end{equation} 

The system (\ref{eqzero}) is a linear homogeneous system with respect to the coefficients $C_{13}$ and $C_{33}$. A necessary condition for the existence of a nontrivial solution is the zero value of its determinant, $\Delta(\tilde{E})=0$. This condition gives us an equation for the complex energy $\tilde{E}=Re E-i \Gamma$ of a quasistationary state which solution can be obtained numerically. The magnitude $\Gamma$ determines the inverse lifetime of the quasistationary state. A natural unit for $\Gamma$ is given by the model parameters as 

\begin{equation}
\Gamma_0=\frac{A}{2L}.
\label{gamma0}
\end{equation}

For a typical example with the QD width $L=40$ nm one has $\Gamma_0=4.5$ meV.
Since $\Gamma$ usually depends exponentially on the barrier width $d$, it is natural to consider the dependence of $\rm{ln} (\Gamma/\Gamma_0)$ on $d$. Such dependence being equal for both of two discrete levels from the described basic example is shown as a function of the barrier width $d$ by the dash-dotted line A in Fig.2(a). As  expected, the lifetime $\tau=\hbar/2 \Gamma$ scales exponentially with the barrier width $d$. For a sufficiently wide barrier with $d \ge 100$ nm one obtains $\Gamma=(5.4 \cdot 10^{-4} \ldots 1.1 \cdot 10^{-5})$ meV which corresponds to the lifetime $\tau=0.6 \ldots 29$ ns. According to our estimates of the manipulation time for the level population produced by the periodic electric field \cite{jetp2020} the required time intervals are of the order of $(4 \ldots 11)$ ps. By comparing it with the obtained lifetimes one can conclude that the safety gap of almost three orders of magnitude is available for the quasistationary states to reside inside the QD before the leakage becomes significant.

\begin{figure}[ht]
\includegraphics[width=105mm]{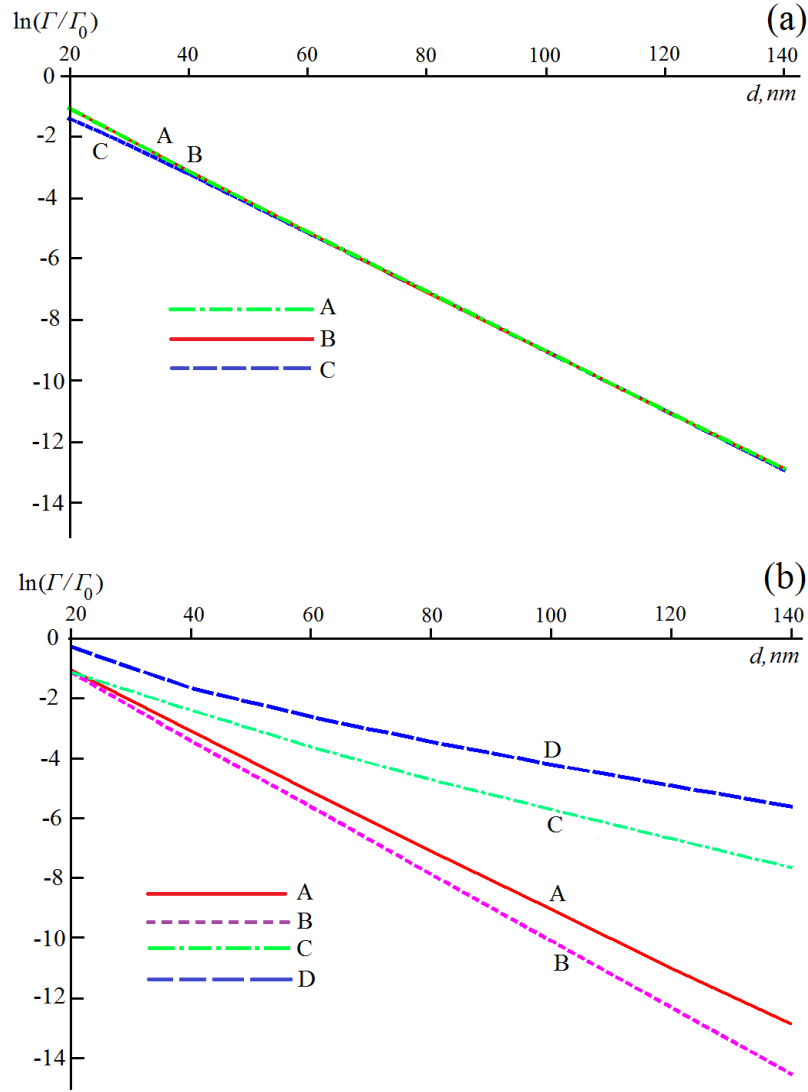}\hspace{1.3pc}%
\begin{minipage}[b]{50mm}\caption{\label{figfit} (a) Dependence for the logarithm of dimensionless level broadening $\Gamma / \Gamma_0$ for the levels $E_{1,2}=\mp 9.62$ meV on the barrier width $d$ in a double barrier structure with identical barriers having the magnitude $M=20$ meV and parallel magnetization $\theta=0$. Line A is for the quasistationary level width obtained during the solution of system (\ref{eqzero}), line B is for the double barrier resonant scattering approach (\ref{te}), line C is for the approximation (\ref{gamma1b}). (b) Same is for the different barrier configurations obtained within the approach (\ref{te}): line A is for the identical barriers with parallel magnetization shown in panel (a), lines B and C are for the barriers with the same height but with antiparallel magnetization $\theta_1=0$, $\theta_2=\pi$, corresponding to the medium level $E_0=0$ and lower (upper) levels $E_{1,2}=\mp 18.1$ meV, and line D is for the structure with parallel but asymmetrical barriers having the magnitudes $M_1=20$ meV and $M_2=10$ meV.}
\end{minipage}
\end{figure}

\section{Quasistationary states lifetime from the scattering problem}

Another way of finding the quasistationary state lifetime is to solve a scattering problem for the double barrier structure \cite{Azbel,Dymnikov,DVbook}. If a plane wave coming from the left in region 1 in Fig.1 is scattered on a double barrier structure, the amplitude of the outcoming wave in region 5 gives the energy-dependent transmission coefficient:

\begin{equation}
T(E)=\frac{4 |D_1 D_2|^2}{\left(|D_1|^2+|D_2|^2\right)^2}
\frac{1}{1+\left(\frac{E-E_n}{\Gamma}\right)^2}.
\label{te}
\end{equation}

Here $D_{1,2}$ are the transmission coefficients for a single barrier 1 and 2 calculated at the resonance energy $E_n$ corresponding to the discrete level in a quantum dot formed by non-transparent barriers. If the energy dependence $T(E)$ is found from the solution of a scattering problem, one may obtain the level broadening $\Gamma$ corresponding to the half-width of the peaks of the function $T(E)$. The scattering problem is solved in our case in a similar way as the problem of the quasistationary states described in the previous Section. Namely, we take the incoming wave having the form (\ref{wf1}) with $C_{11}=1$ and $C_{21}=0$, and look on the solution for the outcoming wave, in particular, on the coefficient $C_{15}$. By definition, we have $T(E)=|C_{15}(E)|^2$.
In Fig.2(a) we plot the energy dependence of  $\rm{ln} (\Gamma/\Gamma_0)$ by the solid line B for the same level broadening as the dash-dotted line A obtained by the approach from the previous Section. It is clear that both approaches agree well with each other.

The last modification of approach suitable for analytical description of the lifetime of the quasistationary states is the analysis of the transmission coefficient for the double barrier structure in a limit of small transparency for each of the barriers. Namely, if two barriers with low transparency $|D_{1,2}| \ll 1$ form a double barrier structure, then the quasistationary state lifetime can be estimated as \cite{DVbook}

\begin{equation}
\Gamma=\Gamma_0 \frac{L}{\tilde{L}} \left( \frac{|D_1|^2+|D_2|^2}{2} \right),
\label{gamma1b}
\end{equation}

where

\begin{equation}
\tilde{L}=L+\frac{1}{2}\frac{d(\phi_1+\phi_2)}{dk}
\label{Ltilde}
\end{equation}

takes into account the k-dependence of the phase factors $\phi_{1,2}$ for the transmission coefficients $D_{1,2}$. The equations (\ref{gamma1b}) and (\ref{Ltilde}) have been obtained in \cite{DVbook} for the problem of scattering of a conventional charge carrier with finite mass. It can be shown that they can be generalized also for the problem of scattering of the massless edge states described by the Hamiltonian (\ref{ham}). For a single barrier the energy dependence for the transmission coefficient $|D_1(E)|^2$ can be easily found in our model for the under-barrier transmission with $|E| < M$:

\begin{equation}
|D_1(E)|^2=\frac{M^2-E^2}{M^2 \rm{ch}^2(\gamma L)-E^2},
\label{d1}
\end{equation} 

where $\gamma$ is given in (\ref{gamma}). The advantage of this approach compared to the previous one dealing with the exact calculations of the double barrier transmission is that all calculations in Eqs (\ref{gamma1b})-(\ref{d1}) can be performed analytically. The results of this approximation are presented in Fig.2(a) by the dotted line C for the same barrier configuration as for the other lines A and B. One can see that despite different nature of applied approaches they all give very close results except the region of very small barrier width $d$ for the curve C where the assumption $|D_{1,2}| \ll 1$ may not be well justified. 

Our final goal is to compare the results for the quasistationary state lifetime obtained for the other configurations of the barriers. In Fig.2(b) we plot the dependence on the barrier width for  $\rm{ln} (\Gamma/\Gamma_0)$ for various configurations within the approach (\ref{te}). Line A is for the previously discussed configuration of two identical barriers with parallel magnetizations. Lines B and C are for the levels formed by the antiparallel barriers with the same height but with magnetization orientations $\theta_1=0$, $\theta_2=\pi$. Here three levels are formed: the medium level $E_0=0$ (line B) and the lower (upper) levels $E_{1,2}=\mp 18.1$ meV (line C). Line D is for the lowest level formed in the structure with parallel but asymmetrical barriers having the magnitudes $M_1=20$ meV and $M_2=10$ meV. One can see that the longest lifetime is described by the curve B. However, this line represents the lifetime of a quasistationary state formed at the level $E=0$ corresponding to the Dirac point of the initial edge spectrum. The degenerate nature of the Dirac point where two dispersion branches cross may create difficulties to use it in applied problems. The longest lifetime for a level in a regular non-degenerate point of the spectrum is achieved by the parallel magnetization of two identical barriers illustrated by curve A in Fig.2(b). As it has been mentioned earlier, it may reach up to $0.6 \ldots 29$ ns for the barriers with the width of $100 \ldots 140$ nm which seems to be accessible for modern micromagnet technology. As a result, coherent manipulation of the discrete level population in the quantum dot on such time scales seems to be possible \cite{jetp2020} which opens the way for the application of such quantum dots in spintronics.

\section{Conclusions}
We have constructed a model of quasistationary states formed at the 1D edge in 2D topological insulator based on HgTe/CdTe quantum well by the application of magnetic barriers with finite transparency. The lifetime of these states is found as a function of barrier geometrical parameters and orientation of their magnetization by different analytical and numerical methods. It is found that a long enough lifetime for a non-degenerate level can be expected with application of two identical barriers (in terms of their width and height) with parallel magnetizations if their width exceeds the threshold of about $100$ nm.

\section*{Acknowledgements}
The authors are grateful to V.A. Burdov and E.Ya. Sherman for valuable discussions. The work is supported by the Ministry of Higher Education and Science of Russian Federation through the State Assignment No 0729-2020-0058 and by the President of Russian Federation grant for young scientists MK-1719.2020.2.

\end{document}